\documentclass{article}
\usepackage{spconf,amsmath,graphicx,hyperref}
\usepackage{booktabs}
\usepackage{multirow}
\usepackage{tcolorbox}
\tcbuselibrary{listings}
\usepackage{caption}
\usepackage{anyfontsize}
\usepackage{tabularx,booktabs,array,ragged2e}
\newcolumntype{Y}{>{\RaggedRight\arraybackslash}X}
\newcommand{\reasoncell}[2]{\textbf{#1} — #2} 
\usepackage[dvipsnames]{xcolor}
\usepackage{soul}          
\usepackage{soulutf8}      

\definecolor{ContentBG}{HTML}{FFF2B2}   
\definecolor{AcousticBG}{HTML}{CDE9FF}  
\definecolor{PredBG}{HTML}{D4F8D4}      

\newcommand{\contentHL}[1]{{\sethlcolor{ContentBG}\hl{#1}}}
\newcommand{\acousticHL}[1]{{\sethlcolor{AcousticBG}\hl{#1}}}
\newcommand{\predHL}[1]{{\sethlcolor{PredBG}\hl{#1}}}

\title{Reasoning Beyond Majority Vote: An Explainable SpeechLM Framework for Speech Emotion Recognition}
\name{\begin{tabular}{c}Bo-Hao Su\textsuperscript{1}, Hui-Ying Shih\textsuperscript{2}, Jinchuan Tian\textsuperscript{1}, Jiatong Shi\textsuperscript{1}\\ Chi-Chun Lee\textsuperscript{2}, Carlos Busso\textsuperscript{1}, Shinji Watanabe\textsuperscript{1}\end{tabular}}
\address{
\textit{\textsuperscript{1}Language Technologies Institute, Carnegie Mellon University, USA}\\
\textit{\textsuperscript{2}Department of Electrical Engineering, National Tsing Hua University, Taiwan}}
\begin{document}
\ninept

\maketitle
\begin{abstract}
\vspace{-1mm}
Speech Emotion Recognition (SER) is typically trained and evaluated on majority-voted labels, which simplifies benchmarking but masks subjectivity and provides little transparency into \emph{why} predictions are made. This neglects valid minority annotations and limits interpretability. We propose an explainable Speech Language Model (SpeechLM) framework that frames SER as a generative reasoning task. Given an utterance, the model first produces a transcript, then outputs both an emotion label and a concise natural-language rationale grounded in lexical and acoustic cues. 
Rationales are generated by a reasoning-capable teacher LLM and used as intermediate supervision, combined with majority labels during fine-tuning.
Unlike prior work primarily focused on boosting classification accuracy, we aim to enhance \emph{explainability} while preserving competitive performance. To this end, we complement majority-label metrics with annotator-aware scoring that credits matches with any annotator label. On MSP-Podcast v1.12, our model maintains improvements over zero-shot SpeechLM baselines, and produces rationales that human evaluators find plausible and well grounded. This demonstrates that incorporating rationale supervision offers a practical path toward interpretable SER without sacrificing predictive quality.
\end{abstract}

\vspace{-3mm}
\section{Introduction}
\vspace{-2mm}
\label{sec:intro}
Speech Emotion Recognition (SER) has traditionally framed affect inference as a single-label classification problem trained and evaluated on majority-voted annotations. While this convention simplifies modeling and benchmarking, it implicitly assumes a unique “gold” label per utterance and disregards the inherent variability in how emotions are perceived. In practice, emotional perception is subjective and context dependent, where individual factors, such as cultural background, lived experience, and conversational role, systematically shape interpretation~\cite{pham2007emotion,mesquita2003cultural, fleisig2023majority, kadasi2023unveiling}. Even within a single dataset, annotators may disagree substantially, reflecting genuine ambiguity in speech signals rather than noise. Consequently, reducing diverse judgments to a single consensus label discards valuable information about emotional nuance and fails to capture the range of human perspectives. As a result, purely label-only supervision offers little insight into \emph{why} a model selected a category, limiting interpretability, trustworthiness, and downstream utility in sensitive applications.

These limitations are most acute at \emph{evaluation} time. Majority-only metrics penalize predictions that agree with well-supported but non-majority annotations, thereby underestimating performance on inherently ambiguous utterances and masking where models align with valid minority judgments~\cite{palotti2023analysis}. To address this, SER research has increasingly embraced \emph{soft-label} formulations, where the full annotator distribution or uncertainty-aware targets are preserved in training and evaluation~\cite{wu2024modelling, chou2024embracing, 10888198}. 
Such methods not only yield higher accuracy on ambiguous cases but also align more faithfully with human variability. Beyond scores, a model requires understanding \emph{how} systems map linguistic content and acoustic cues (prosody, pitch, energy, timing) to emotions so we can target weaknesses (e.g., lexical shortcuts vs.\ prosodic evidence). While recent SpeechLLM studies have begun entering the SER domain~\cite{mai2025aa,wang2024blsp,chang2024exploring,bukhari2024selm, li2025revise}, and some report headline accuracies in technical reports~\cite{goel2025audio}, the majority still treat these models primarily as classifiers. Though Chen et al.~\cite{chen25i_interspeech} improve explainable SER via fine-grained speech emotion descriptors, descriptors are predefined from the pretrained model. This paradigm under-utilizes their generative capabilities, leaving untapped the potential to combine reasoning, explanation, and label uncertainty into a more holistic framework for emotion recognition.

In this work, we introduce an explainable SpeechLM framework that outputs a natural-language \emph{rationale} alongside the emotion label. A reasoning-capable teacher LLM synthesizes concise explanations grounded in both lexical content and acoustic cues simultaneously. These rationales serve as intermediate supervision signals for a SpeechLM backbone. This teacher–student rationale supervision is shown effective in NLP that distilling teacher-generated reasoning traces can transfer rationalization capabilities to student models \cite{li2023symbolic}. Given an utterance, the model first emits an ASR transcript to preserve transcription skill, then an emotion decision followed by a brief evidence-based rationale, all within a single coherent response. 
This structure encourages the model to provide both the emotion prediction and its rationale, improving interpretability while leaving majority-label training unchanged. 
To our knowledge, this is the first SER study to pair qualitative/quantitative rationale analysis with annotator-aware evaluation on a large in-the-wild corpus. 


Empirically, our approach improves interpretability while maintaining competitive accuracy against traditional SER classifiers. Compared with zero-shot explainable CoT baselines, our model consistently performs better. On the MSP-Podcast v1.12 \cite{lotfian2017building} test set, the model achieves higher Macro-F1 under conventional majority-label metrics, and the advantages persist when scoring on the union set of all annotators' labels. Qualitative analyses show that the generated rationales frequently highlight salient acoustic–linguistic cues and are broadly accepted by humans. Moreover, external evaluation with \emph{LLMs-as-judges} confirms that our rationales are consistently preferred over those from strong zero-shot baselines. In summary, our contributions are threefold: (i) an explainable SER framework that jointly produces labels and rationales, (ii) an evaluation protocol combining majority-label and annotator-aware Macro-F1 while keeping training scheme unchanged with rationales augmented, and (iii) empirical evidence showing that rationale-augmented SpeechLMs better align with both human and LLM-based judgments.

\vspace{-3mm}
\section{Methodology}
\label{sec:method}
\vspace{-3mm}

We propose an explainable SpeechLM that produces a natural-language rationale with the final emotion decision at the same time. The framework comprises two components: (i) \emph{reasoning generation} by a multimodal teacher LLM, and (ii) \emph{supervised fine-tuning} of a backbone SpeechLM using these rationales as intermediate supervision. We detail the reasoning module below.
\vspace{-3mm}
\subsection{Reasoning Generation}
\label{sec:reasoning_generation}
\vspace{-2mm}
Recent advances in reasoning-capable LLMs have made strong open models widely available. To bridge the gap between speech inputs and categorical emotion outputs, we leverage these models to generate human-readable rationales for each input–output pair in speech emotion recognition (SER). Note that rationales are conditioned on the gold label and thus reflect a label-informed explanation. The prompt template used for rationale synthesis is provided in the \emph{Prompt Design} box, where the \$CONTENT and \$EMOTION mean the transcript and ground-truth emotion labels of the training utterances respectively.
\vspace{-1mm}
\begin{tcolorbox}[title=Prompt for Reasoning Generation, colback=gray!5, colframe=gray!40!black, sharp corners=south, listing only, listing options={basicstyle=\ttfamily\small, breaklines=true}]
\vspace{-2mm}
Given the Content: \{\$CONTENT\}

Emotion: \{\$EMOTION\}

Provide a detailed reasoning that explains how the emotion and its intensity relate to the spoken content and the audio's characteristics, such as tone, pitch, speaking style, clarity, and other acoustic properties.

\#\# Here are some rules you should follow:

- DO NOT use the word \{\$EMOTION\} in your reasoning.

- DO NOT use open-ended questions or follow-up invitations in the reasoning. For example, “What do you think about this?", “What do you think?" or “If you want to talk more about this or have other thoughts, feel free to share."

- DO NOT use “Well, you know" or “Well" in the reasoning.
\vspace{-4mm}
\end{tcolorbox}

Concretely, we adopt \texttt{Qwen2.5-Omni} as the reasoning agent due to its strong multimodal understanding and reasoning capabilities. For each utterance, we provide the audio signal (encoded into embeddings by audio encoder) concatenated with prompt token embeddings consisting of its transcript and the reference emotion label to generate rationales for training data.
Note that unlike prior work, we do not discretize acoustic cues into fixed levels.
This strategy enables diverse natural-language descriptions and reflects real-world settings where the target distribution and class granularity are unknown. All rationales are generated exclusively on the training split, ensuring no label information leaks into validation or test evaluation.

To preserve explainability, we impose the following constraints: (1) the rationale must \emph{not} state or paraphrase the gold label; (2) it must integrate both content and acoustic considerations rather than relying on transcript alone; and (3) it must avoid trailing open-ended prompts or meta-commentary (e.g., “What do you think?”). We apply simple post-processing to strip disallowed phrases and discard generations that violate length or evidence-coverage requirements. The resulting rationales ($r$) and corresponding emotion labels ($e$) serve as intermediate supervision for our backbone model in the subsequent fine-tuning stage.
\vspace{-2mm}
\subsection{Supervised Fine-Tuning (SFT)}
Given the curated triples $\{(x_i, e_i, r_i)\}_{i=1}^N$ consisting of speech $x$, gold emotion $e$ and rationale $r$ from previous stage, we fine-tune the backbone SpeechLM (\texttt{Qwen2Audio}) with Low-Rank Adaptation (LoRA)~\cite{hu2022lora}. We adapt only the query, key, value and output projection (Q/K/V/O) weights in self-attention blocks and the language-model head, keeping the remaining parameters frozen. To mitigate catastrophic forgetting and overfitting, we employ a small learning rate with warmup and cosine decay, and weight decay on LoRA parameters.

In the target formatting, each training instance is rendered as a single autoregressive target string to preserve ASR ability with reference transcript $a_i$ while supervising the final decision and reasoning:
\begin{align}
y_i &= \text{\texttt{<asr>}}\, {a}_i\, \text{\texttt{</asr>}} \nonumber \\[-0.2ex]
& \text{\texttt{<answer>}}\, \text{I think the emotion is } e_i\, \text{\texttt{</answer>}} \nonumber \\[-0.2ex]
& \text{\texttt{<reason>}}\, r_i\, \text{\texttt{</reason>}}
\label{eq:string},
\end{align}
where $e_i$ is a single emotion token from the closed set.

Therefore, let \(x_i\) be the input audio, and \(\mathbf{y}_i=(y_{i,1},\dots,y_{i,t})\) the tokens, where $t$ stands for time steps.
The SFT objective is:
\begin{equation}
\mathcal{L}_{\mathrm{SFT}}(\theta)
= -\frac{1}{B}
\sum_{i}\;\sum_{t}
\log p_\theta\!\big(y_{i,t}\,\big|\,\mathbf{y}_{i,<t},\,x_i\big).  
\label{eq:loss}
\end{equation}
Here \(\theta\) denotes the LoRA-augmented model parameters, and $B$ represents the batch size. Only LoRA adapters are updated. The complete dataset composition and fine-tuning configurations are described in Section~\ref{sec:experiment_setup}.

\begin{table*}[htb]
    \centering
    \caption{Baseline model comparison with Macro-F1 score on the MSP-PODCAST testing set
    ; $^\star$ denotes p-value $\leq$ 0.05.}
    \scalebox{0.9}{
    \begin{tabular}{cc|c|c|c|c|c|c|c|c}
        \toprule
               & & \multicolumn{4}{c|}{Open-form} & \multicolumn{4}{c}{Closed-form}\\
              \midrule
              & & \multicolumn{2}{c|}{All Testing Set} & \multicolumn{2}{c|}{agreement $>$ 0.5} & \multicolumn{2}{c|}{All Testing Set} & \multicolumn{2}{c}{agreement $>$ 0.5} \\
        \midrule
        id & Models &  MV-labels & All-labels & MV-labels & All-labels & MV-labels & All-labels & MV-labels & All-labels\\
        \midrule
        1 & WavLM~\cite{Busso2025mspodcast} & - & - & - & - & \underline{29.70\%} & - & - & - \\
        2 & Wav2vec 2.0~\cite{Busso2025mspodcast} & - & - & - & - & 23.80\% & - & - & - \\
        3 & HuBERT~\cite{Busso2025mspodcast} & - & - & - & - & 28.50\% & - & - & - \\
        \hline
        4 & SALMONN & 16.91\% & 33.26\% & 19.21\% & 31.52\% & 12.32\% & 20.95\% & 13.90\% & 20.11\%\\
        5 & Q2A & 19.53\% & 39.49\% & 20.93\% & 34.61\% & 22.96\% & 45.16\% & 24.66\% & 40.24\%\\
        6 & Q2A+\textit{CoT} & \underline{27.80\%} & \underline{52.42\%} & \underline{30.09\%} & \underline{46.87\%} & 26.39\% & 49.56\% & 28.60\% & 44.79\% \\
        7 & Baichuan-Omni-1.5 & 21.51\% & 46.27\% & 22.60\% & 40.24\% & 21.52\% & 46.16\% & 22.63\% & 40.38\% \\
        8 & Q2.5O & 24.54\% & 48.12\% & 26.49\% & 43.36\% & 23.31\% & 47.17\% & 25.44\% & 42.27\% \\
        9 & Q2.5O+\textit{CoT} & 26.00\% & 49.41\% & 28.09\% & 44.61\% & 28.70\% & \underline{57.37\%} & \underline{31.81\%} & \underline{52.74\%} \\
        \midrule
        \midrule
        10 & Q2A-lora-SFT w/o reasoning & 30.11\% & 56.78\% & 32.15\% & 50.30\% & 26.76\% & 49.35\% & 31.31\% & 43.81\% \\
        11 & Q2A-lora-SFT w/ reasoning & $\textbf{36.14\%}^{\star}$ & $\textbf{66.78\%}^{\star}$ & $\textbf{40.15\%}^{\star}$ & $\textbf{59.32\%}^{\star}$ & $\textbf{30.59\%}^{\star}$ & $\textbf{62.55\%}^{\star}$ & $\textbf{33.64\%}^{\star}$ & $\textbf{54.13\%}^{\star}$ \\
        \bottomrule
    \end{tabular}
    }
    \vspace{-3mm}
    \label{tab:baseline_comparison}
\end{table*}

\vspace{-3mm}
\section{Experiment}
\label{sec:exp}
\vspace{-2mm}
\subsection{Experiment Setup}
\label{sec:experiment_setup}
We construct the training set by pooling CREMA\textendash D~\cite{cao2014crema}, IEMOCAP~\cite{busso2008iemocap}, MELD (train)~\cite{poria2019meld}, RAVDESS~\cite{livingstone2018ryerson}, and MSP\textendash Podcast v1.12 (train), yielding approximately $132$ hours of SFT data with paired speech–label-rationale triples. All emotion categories present in the training corpora are already covered within the eight categories defined in MSP-Podcast, so no additional label mapping is required. For evaluation, we use the official MSP\textendash Podcast v1.12 Test\_1 split due to its scale and diversity.

All experiments were run with LLaMA-Factory~\cite{zheng2024llamafactory} in \texttt{sft} mode on \texttt{Qwen2-Audio-7B-Instruct} using LoRA (rank $=8$, $\alpha=16$). We fine-tune for $2$ epoch (decided by early stopping) with learning rate $1{\times}10^{-5}$, cosine scheduler with $10\%$ warmup, \texttt{bf16} precision, per-device batch size $=1$, and gradient accumulation $=8$. We cap the training set at \texttt{max\_samples}$=15,000$ and use a maximum sequence length of $2,048$ tokens. Data are formatted with the \texttt{qwen2\_audio} template so that only the assistant segment (containing transcript $a_i$, emotion $r_i$, and the rationale $r_i$ in Eq.~\eqref{eq:string}) contributes to the token-level cross-entropy loss in Eq.~\eqref{eq:loss}.

For evaluation, we compute the Macro-F1 score for the overall performance on the eight-category emotion prediction task excluding others in the MSP-Podcast. In this work, to further capture alignment with minority judgments, we include both majority emotion (\textit{MV-labels}) and annotator-aware Macro-F1 (\textit{All-labels}) in our final reports. In annotator-aware Macro-F1, counting a prediction as correct if and only if $\hat y_i \in A_i$, where $A_i$ is the union set of all annotators' labels (i.e., it matches \emph{any} annotator’s label for that item).
We also report results on the strict-majority subset—items with one label receiving $>$50\% of votes (distinct from plurality).

Given outputs comprising an open-form rationale and a closed-form decision. Open-form means that we do not limit the output space of the prediction to a subset of emotional classes. In the open-form evaluation, we simply adopt the prompt \texttt{"What emotion do you think the audio expresses?"} with audio as input. Otherwise, we call it closed-form.
The inputs of the closed-form are the audio and prompt:

\texttt{"What emotion do you think the audio \\expresses? Choose one from the following \\options:$\backslash n$A. Angry$\backslash n$B. Sad$\backslash n$C. Happy$\backslash n$\\D. Surprise$\backslash n$E. Fear$\backslash n$F. Disgust$\backslash n$\\G. Contempt$\backslash n$H. Neutral$\backslash n$I. Other"}.

Then, we employ \texttt{GPT-OSS-20B} as a parser to extract the predicted emotion. We subsequently prompt \texttt{ChatGPT-5} to cluster labels into nine categories—eight target emotions and \emph{other}, which includes prediction failures and any label outside the target set. We provide all detailed prompts for extraction and grouping on our demo page\footnote{\href{https://subohao.github.io/ICASSP26-Emotion-Reasoning/}{Demo page}} to enhance the reproducibility of our experiments.

\vspace{-3mm}
\subsection{Baseline Comparisons}
\vspace{-2mm}
In this work, we include open-source baseline SpeechLMs as our baseline models, including Qwen2-Audio-Instruct-7B (Q2A)~\cite{chu2024qwen2}, Qwen2.5-Omni (Q2.5O)~\cite{xu2025qwen2}, SALMONN~\cite{tang2023salmonn} and Baichuan-Omni-1.5~\cite{li2025baichuan}. Besides, we list the most recent result~\cite{Busso2025mspodcast} from the MSP-Podcast Corpus as traditional classification baseline models utilizing self-supervised representation (WavLM, Wav2vec 2.0, and HuBERT) as referenced SFT point. For ease of reference, Table~\ref{tab:baseline_comparison} includes a model-index (id) column that we use to cross-reference systems in the results discussion.


Table~\ref{tab:baseline_comparison} summarizes results on the MSP-Podcast test set across open-form and closed-form prediction. Model 4, 5, 7, and 8 represent strong zero-shot baselines. Surprisingly, closed-form Macro-F1 is inconsistent across models and generally lower than open-form, likely because emotions are subtle and benefit from fine-grained wording: open-form lets the LLM leverage broad lexical knowledge, whereas closed-form’s fixed label tokens restrict precision and accentuate label-prior and tokenization errors.

We additionally report the effect of Chain-of-Thought (\textit{CoT}) prompting.
As expected, \textit{CoT} improves most systems by encouraging more structured reasoning, though the overall accuracy remains modest. For instance, Q2A+\textit{CoT} achieves 27.8\% Macro-F1 (MV-labels, open-form), while Q2.5O+\textit{CoT} improves to 26.0\%, indicating that CoT is a useful but limited tool in the zero-shot setting. Interestingly, Q2A+\textit{CoT} slightly outperforms Q2.5O+\textit{CoT} in SER. This result highlights the trade-off between specialization and generalization. Q2A, being an audio-specialized model, is more sensitive to fine-grained prosodic cues such as pitch, energy, and rhythm, which are critical for distinguishing emotional states. Despite the versatility of multimodal generalist models, this finding suggests that task-specialized audio models retain a marginal advantage in domains where paralinguistic features are paramount.

\begin{figure}
    \centering
    \includegraphics[width=\linewidth]
    {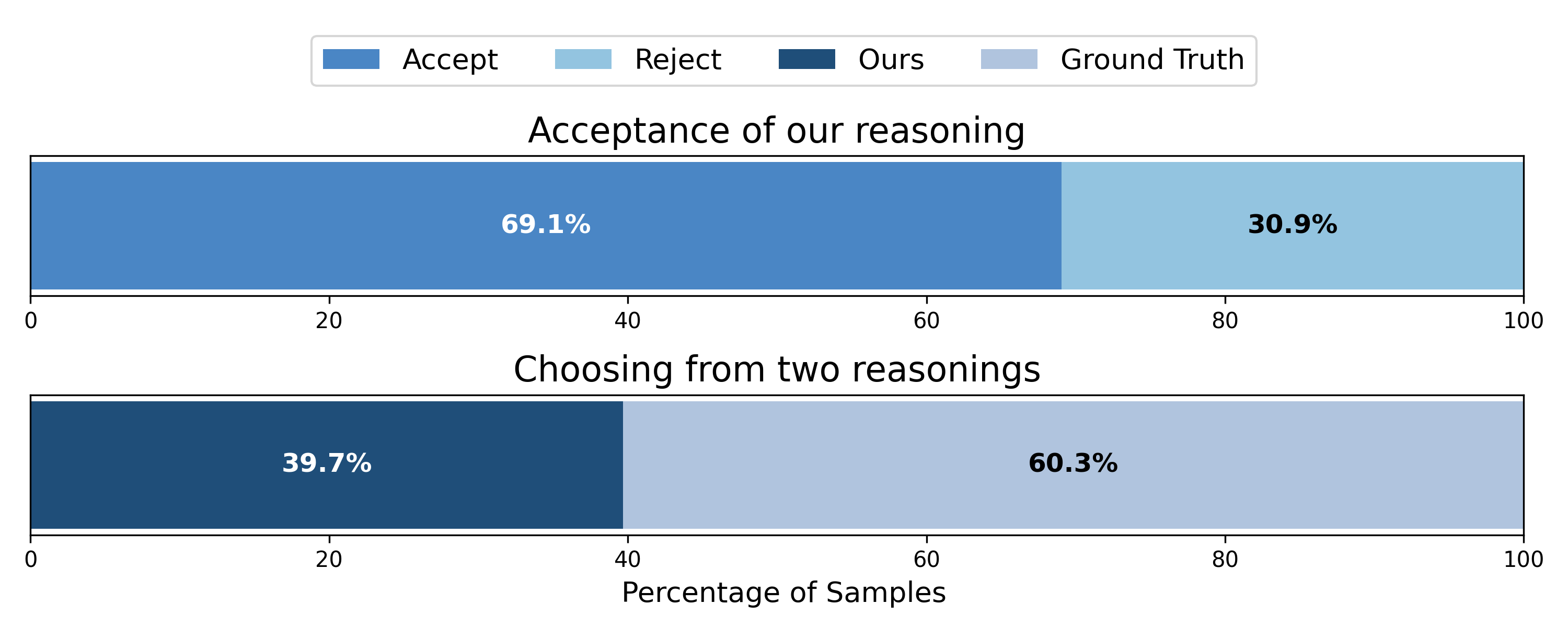}
    \caption{Subjective evaluation results (10 raters).}
    \label{fig:sub_fig}
    \vspace{-6mm}
\end{figure}


\begin{table*}[htb]
\small
\centering
\setlength{\tabcolsep}{4pt}
\renewcommand{\arraystretch}{1.12}
\caption{Open-form reasoning demonstrations on sample utterances. Each cell shows label and the model’s rationale, where the \contentHL{yellow}, \predHL{green} and \acousticHL{blue} color indicate the rationale corresponding to content, prediction and acoustic information, respectively.}
\label{tab:reasoning_examples_open}
\scalebox{0.8}{
\begin{tabularx}{\textwidth}{@{}l l Y @{}}
\toprule
\textbf{ID} & \textbf{MV-labels} & \textbf{Ours} \\
\midrule
0340\_0148 & Surprise 
& \reasoncell{Surprise}{When someone says \contentHL{“I didn't know that she has pink hair. How did I miss that?"} it sounds like they're a bit surprised. \predHL{I think the emotion is surprise.} \acousticHL{The tone might be a bit raised, a bit more excited than usual. And the pitch could be a bit higher too.} \contentHL{It's like they're realizing something new and a bit shocked.}} \\
3737\_0192 & Angry 
& \reasoncell{Angry}{The speaker is saying \contentHL{“fight with me, like, oh, again, again, again, you forgot our anniversary, if -".} \predHL{I think the emotion is angry.} \acousticHL{The tone seems to be a bit frustrated or annoyed. The pitch might be a bit higher, indicating some level of irritation. The speaking style is a bit repetitive, which could show that the speaker is trying to emphasize the importance of the anniversary. The clarity of the speech is good, so it's easy to understand the words.} \contentHL{Overall, it seems like the speaker is a bit upset about the anniversary being forgotten.}}\\
2991\_1041 & Neutral & \reasoncell{Sad}{The content \contentHL{“no possibility of saving anything" suggests a sense of hopelessness or despair.} \predHL{I think the emotion is sad.} \acousticHL{The tone of the speaker might be low and slow, indicating a lack of energy or motivation. The pitch could be lower than normal, reflecting a feeling of being overwhelmed or defeated. The speaking style might be hesitant or even shaky, as if the speaker is struggling to find the words to express their feelings. The clarity of the speech could be affected by the speaker's emotional state, possibly making it a bit muddled.} \contentHL{All these factors contribute to the overall sad emotion.}}\\
\bottomrule
\end{tabularx}
}
\vspace{-6mm}
\end{table*}

In contrast, our proposed SFT-based model (Model 11) substantially outperforms all zero-shot baselines (including \textit{CoT}-based models), reaching 36.1\% Macro-F1 on MV-labels (open-form) and 30.6\% in the closed-form setting. Importantly, these gains are not merely a byproduct of fine-tuning: traditional SFT models trained directly for SER (Models 1–3) achieve only 23–30\% on closed-form metrics, highlighting that simply applying supervised training does not guarantee strong performance. Instead, the improvement of Model 11 stems from incorporating rationale supervision during fine-tuning, which enhances both predictive accuracy and interpretability.

We emphasize that the comparison is not entirely symmetric and our model is fine-tuned while baselines are zero-shot. However, by including conventional SFT-trained systems (Models 1–3) in the table, we provide a fairer reference point. Against this stronger baseline, our approach still achieves clear improvements, demonstrating that rationale-guided supervision provides additional benefits beyond standard fine-tuning.

\vspace{-3mm}
\subsection{Ablation on Rationales}
\vspace{-2mm}
To further assess the contribution of rationale supervision, we compare our proposed model trained with reasoning (Model 11) against its counterpart trained without reasoning (Model 10).
Crucially, as described in Section~\ref{sec:reasoning_generation}, adding rationale supervision (Model 11) yields the strongest results, surpassing both Model 10 and all baseline systems. The gains are consistent across open-form and closed-form evaluation and persist under annotator-aware Macro-F1, confirming that rationales not only enhance interpretability but also serve as effective auxiliary training signals. 
This ablation again supports that our performance specifically arise from integrating rationale generation into training.

\vspace{-2mm}
\subsection{Subjective Test}
\vspace{-2mm}
To assess rationale quality, we design two human studies (Test I and Test II), where ten raters participated in each study. All human raters were volunteer lab members, currently pursuing Master’s or Ph.D. degrees.
Raters listened to the audio, read the model’s predicted label and rationale, and provided judgments in the end. Items were randomized and raters were blind to correctness and system identity.
\begin{description}
\vspace{-1mm}
\item[\textbf{Test I - Accept/Reject}:] To assess plausibility of the label-rationale pair, we sampled $32$ utterances: $16$ where the predicted label matches the label of the reference, and $16$ where it does not (balanced by emotion). Raters marked the pair as \emph{Accept} or \emph{Reject}. We report overall Accept rate.
\vspace{-1mm}
\item[\textbf{Test II - Plausibility}:] To assess the reasoning plausibility, we selected $32$ utterances where the model’s predicted label disagreed with the reference. For each utterance, we formed two candidates including \emph{Model-predicted} and \emph{Gold-referenced} (rationale generated by our prompting procedure using the reference label as guidance). Raters marked the preference one, and we report the Gold-vs-Predicted preference rate.
\end{description}
\vspace{-1mm}
Figure~\ref{fig:sub_fig} summarizes the results. In Test I, the overall average acceptance rate reached 69.1\% (95\% confidence intervals(CIs): [58.9\%, 79.2\%]), with 72.5\% for the match condition and 65.6\% for the non-match condition. These results indicate that the reasoning generated by our model is perceived as plausible by human raters, even when the predicted emotion does not align with the ground truth.

In Test II, our rationales were preferred over gold-referenced ones in 39.7\% of cases (95\% CIs: [30.2\%, 49.2\%]). This finding suggests that providing explicit reasoning alongside emotion predictions can sometimes outweigh strict alignment with majority-vote labels. In other words, human raters may value the plausibility and interpretability of the reasoning process itself, highlighting the importance of considering minority labels as correct under our proposed explainable modeling framework.

Table~\ref{tab:reasoning_examples_open} presents open-form reasoning examples from our model. In the third example of Table~\ref{tab:reasoning_examples_open}, although the majority-vote label is neutral (3 neutral and 2 sad among all annotations), after reviewing the reasoning and predictions, 9 raters judged our model’s output as more plausible and better aligned with the audio.
The rationales explicitly cite linguistic content and acoustic/prosodic cues, providing evidence for the predicted label and improving interpretability. Additional examples and side-by-side comparisons are available in our demo page.

        


\begin{figure}
    \centering
    \includegraphics[width=\linewidth]{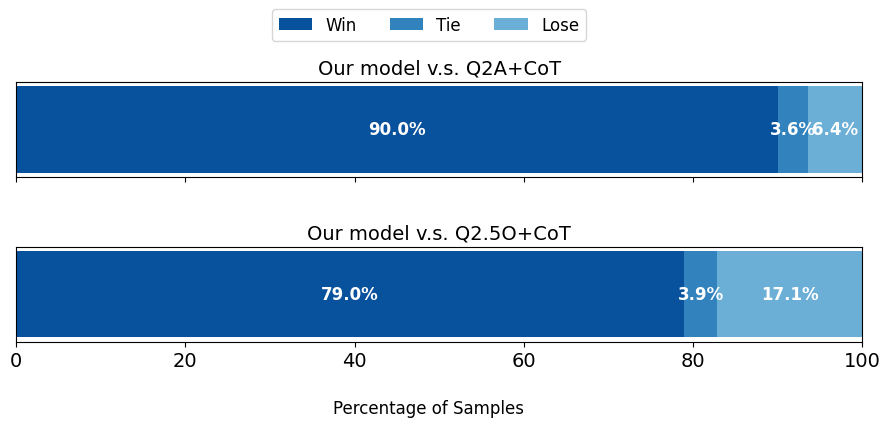}
    \caption{Win-rate comparison by gemini-2.0-flash}
    \vspace{-6mm}
    \label{fig:win-rate_fig}
\end{figure}


\vspace{-3mm}
\subsection{LLMs as judges}
\vspace{-2mm}
Recent work increasingly uses \emph{LLMs-as-judges} to evaluate open-ended model outputs, showing strong agreement with human preferences by using strong LLMs~\cite{dong2024can, zheng2023judging, chen2025audio}.
In this evaluation process, we conduct win-rate comparison by using gemini-2.0-flash API between the reasoning generated through our proposed model, Q2A+\textit{CoT}, and Q2.5O+\textit{CoT}. To fairly compare these samples, we select samples that deliver the same final prediction from comparison models, and then we randomly sample 50 items from each emotion category. To avoid LLMs preferring our model, we do not provide the model information of the input reason by simply mentioning model A and B. 
The overall results are presented in Fig~\ref{fig:win-rate_fig}, where our model’s reasoning was preferred in 90.0\% (95\% CIs: [87.5\%, 92.4\%]) of pairs vs. Q2A+\textit{CoT} and 79.0\% (95\% CIs: [75.7\%, 82.2\%]) vs. Q2.5O+\textit{CoT}. These results confirm that rationale supervision not only improves predictive accuracy but also produces explanations that are consistently judged as more convincing and informative by a strong commercial LLM evaluator. Detailed win-rate prompt could be also found from our demo page.





    







    
    
    


\vspace{-3mm}
\section{Conclusions}
\label{sec:majhead}
\vspace{-2mm}
We proposed an explainable SpeechLM that produces transcripts, rationales, and emotion labels in a unified response. Trained with majority labels and teacher-synthesized rationales, the model achieves competitive accuracy while improving interpretability. On MSP-Podcast v1.12, it outperforms zero-shot baselines under both majority-label and annotator-aware Macro-F1. Human studies and LLM-as-judge evaluations further confirm the plausibility and preference of its explanations. These findings show that rationale supervision offers a practical path to transparent SER without sacrificing accuracy.
Future work includes zero-shot extensions, conversational corpora, and explicit uncertainty modeling.

\vspace{-2mm}
\section{Acknowledgment}
\label{sec:typestyle}
\vspace{-1mm}
Experiments of this work used the Bridges2 system at PSC
and Delta system at NCSA through allocations CIS210014
and IRI120008P from the Advanced Cyberinfrastructure Coordination Ecosystem: Services $\&$ Support (ACCESS) program, supported by National Science Foundation grants
\#2138259, \#:2138286, \#:2138307, \#:2137603, and \#:2138296. Thanks to the National Science and Technology Council(NSTC) for the funding under 114-2917-I-564-022.

\begingroup
\fontsize{8.pt}{9.pt}\selectfont  
\bibliographystyle{IEEEbib}
\bibliography{strings,main}
\endgroup

\end{document}